\documentclass[fleqn,twoside]{article}
\usepackage[headings]{espcrc2}

\readRCS
$Id: espcrc2.tex,v 1.2 2004/02/24 11:22:11 spepping Exp $
\usepackage{graphicx}
\newcommand{\psla}{p\kern-.45em/}
\newcommand{\sq}{\ifmmode{\tilde{q}} \else{$\tilde{q}$} \fi}
\newcommand{\sg}{\ifmmode{\tilde{g}} \else{$\tilde{g}$} \fi}
\newcommand{\beq}{\begin{equation}}
\newcommand{\eeq}{\end{equation}}
\newcommand{\bea}{\begin{eqnarray}}
\newcommand{\eea}{\end{eqnarray}}

\newcommand{\AmS}{{\protect\the\textfont2
  A\kern-.1667em\lower.5ex\hbox{M}\kern-.125emS}}

\title{Two-loop SUSY QCD correction to the gluino pole mass }

\author{Youichi Yamada\address{Department of Physics, Tohoku University, 
Sendai 980-8578, Japan }   
       \thanks{Talk at the 7th International Symposium 
       on Radiative Corrections (RADCOR2005), Shonan Village, 
      Japan, Oct. 2-7, 2005} 
}
       
\begin{document}

\begin{abstract}
We calculate the pole mass of the gluino as a function of the running 
parameters in the lagrangian, to $O(\alpha_s^2)$ in SUSY QCD. 
The correction shifts the pole mass from the running mass 
by typically 1--2 \%. 
This shift can be larger than the expected accuracy 
of the mass determination at future colliders, and 
should be taken into account for precision studies of the SUSY breaking 
parameters. 
The effects of other corrections are breifly commented. 
\vspace{1pc}

TU-763
\end{abstract}

\maketitle

\section{Introduction}

Extention of the standard model by supersymmetry (SUSY), 
with the breaking scale not much higher than the electroweak scale, 
has been studied as a very promising solution to the 
hierarchy problem between the electroweak scale and 
the Planck/grand unification scale. In these models, 
such as the minimal supersymmetric standard model (MSSM) \cite{mssm}, 
all particles in the standard model have their superpartners 
with masses below $O(1)$ TeV. 
These new particles are then expected 
to be produced at colliders in near future, 
such as the CERN Large Hadron Collider (LHC) 
and the International Linear Collider (ILC). 

One of the main motivations for the experimental study of these 
new particles \cite{lhcilc}, the SUSY particles, is the determination 
of the soft SUSY breaking parameters \cite{msusydeterm,SPA}, 
which gives an important information of 
the SUSY breaking mechanism in the unified theory. 
For example, the unification of three gaugino masses 
$(M_3,M_2,M_1)$ at the same scale as that of the gauge couplings 
is a crucial test for the SUSY grand unified 
theory \cite{susygut1,susygut2} and 
superstring phenomenology \cite{string}. 

For this purpose, in addition to the precise measurements of 
the physical parameters of the SUSY particles \cite{lhcilc}, 
one also needs precise prediction of the relations 
between these observables and parameters in the lagrangian. 
In some cases, we have to calculate the relations beyond 
the one-loop order to match expected experimental precision. 
For example, the two-loop mass corrections have been calculated 
for the top and bottom quarks \cite{mq2loopsusy}, 
squarks in the first two generations \cite{msq2loop}, 
and Higgs bosons \cite{mh2loop}. 

Here we focus our attention to the mass of the gluino $\tilde{g}$, 
the SU(3) gaugino, in the framework of the MSSM. 
At the LHC, gluino is, if it is sufficiently light, 
expected to be copiously produced \cite{sgprod}. 
Its mass $m_{\sg}$ can be determined from the distributions of 
the decay products. A study \cite{msgdet1} shows that, 
for the SUSY parameter set SPS1a given in Ref.~\cite{SPS1a} 
with $m_{\sg}\simeq 600$ GeV, $m_{\sg}$ can be determined 
to accuracy $\delta m_{\tilde{g}}=8$ GeV from 
precision data at the LHC with $300$ fb$^{-1}$, and 
even to $\delta m_{\tilde{g}}=6.5$ GeV 
when combined with the data from the ILC. 
% [ LC($\sim 100-1000$ fb$^{-1}$, to 1 TeV)] data 
On the other hand, the one-loop QCD contribution to the 
the difference between the pole mass $m_{\sg}$ and the running 
mass $M_3$ of the gluino \cite{mv1,pierce1,pierce2} is much larger, 
typically $O(10)$ \%. 
One therefore naively expect the two-loop mass correction 
might be $O(1)$ \%, similar to the experimental uncertainty. 
It is therefore important to examine whether higher-order corrections 
to the gluino mass is really relevant in the determination of the 
SUSY breaking parameters at future precision measurements. 

In this talk we present the pole mass of the gluino 
as a function of the lagrangian parameters, 
including $O(\alpha_s^2)$ SUSY QCD correction 
obtained by diagrammatic calculation \cite{myletter}. 

\section{Two-loop SUSY QCD mass correction}

The pole mass $m_{\sg}$ of the gluino, which is defined by  
the complex pole $s_p=(m_{\sg}-i\Gamma_{\sg}/2)^2$ of 
the gluino propagator, is given at the two-loop order as 
\begin{equation}
m_{\sg} = M_3 + \delta m_{\sg}^{(1)} +\delta m_{\sg}^{(2)}, 
\end{equation}
where the corrections $\delta m_{\sg}^{(1,2)}$ are expressed 
in terms of the one-loop and two-loop parts of the gluino self energy 
%\begin{equation}
$\Sigma(p) = \Sigma_K(p^2)\psla + \Sigma_M(p^2)$ 
%\end{equation}
as 
\begin{eqnarray}
\delta m_{\sg}^{(1)} 
&=& - {\rm Re}[ M_3\Sigma_K^{(1)}(M_3^2)+\Sigma_M^{(1)}(M_3^2) ], 
\nonumber\\ 
\delta m_{\sg}^{(2)} 
&=& - {\rm Re}[ M_3\Sigma_K^{(2)}(M_3^2)+\Sigma_M^{(2)}(M_3^2) ]
\nonumber \\ 
&& + {\rm Re} \left[ 
\{ M_3\Sigma_K^{(1)}(M_3^2)+\Sigma_M^{(1)}(M_3^2) \}
\right. 
\nonumber\\ 
&&  
\times \{ \Sigma_K^{(1)}(M_3^2) + 2M_3^2\dot{\Sigma}_K^{(1)}(M_3^2) 
\nonumber\\ 
&& 
\left.
+ 2M_3\dot{\Sigma}_M^{(1)}(M_3^2) \} \right] . 
\label{pole} 
\end{eqnarray}
Here $M_3$ is the running tree-level gluino mass. 
The dot in Eq.~(\ref{pole}) denotes the derivative with respect to 
the external momentum squared $p^2$. 

The SUSY QCD contribution to $\Sigma(p)$ is generated by loops with 
the gluino, gluon, quarks, and squarks. 
Masses and couplings in the lagrangian 
are renormalized in the $\overline{\rm DR}'$ scheme \cite{drbarprime} 
at the scale $Q$. 
Here we ignore SU(2)$\times$U(1) breaking effects 
in the loops, such as the quark masses and 
squark left-right mixings. This approximation is valid 
for the case where the gluino and squarks are sufficiently 
heavier than the quarks. 
Later we will briefly comment on the effects of these 
SU(2)$\times$U(1) breakings. 
For simplicity, we also assume degenerate mass $m_{\sq}$ for squarks. 

The one-loop correction $\delta m_{\sg}^{(1)}$ 
in our approximation is \cite{pierce1,pierce2} 
\bea
\delta m_{\sg}^{(1)} &=& 
\frac{C_V \alpha_s }{4\pi} M_3 \left( 5 -6\log \frac{M_3}{Q} \right) 
\nonumber \\
&& +\frac{\alpha_s }{\pi} N_q T_F 
M_3 B_1(M_3^2, 0, m_{\sq}) ,
\label{oneloop}
\eea
where $C_V=3$, $T_F=1/2$, and $N_q=6$ is the number of quarks. 
Parameters ($\alpha_s$, $M_3$, $m_{\sq_i}$) 
in Eq. (\ref{oneloop}) are the $\overline{\rm DR}'$ running ones at the 
renormalization scale $Q$. 
$B_1(p^2,m_1,m_2)$ is the one-loop function \cite{PV} in 
the convention of Ref.~\cite{denner}. 

\begin{figure}[ht]
%\begin{figure}[htbp]
%\begin{center}
\includegraphics[width=7cm]{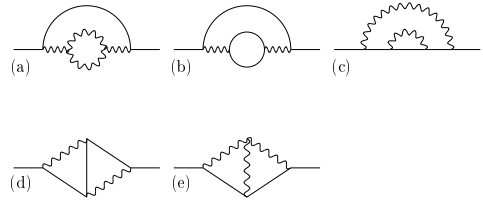}
%\end{center}
\caption{ %\baselineskip=23pt
Two-loop $O(\alpha_s^2)$ contributions to $\delta m_{\sg}^{(2,1)}$, 
without quark and squark propagators. 
}
\label{fig1}
\end{figure}

\begin{figure}[ht]
%\begin{figure}[htbp]
%\begin{center}
\includegraphics[width=8cm]{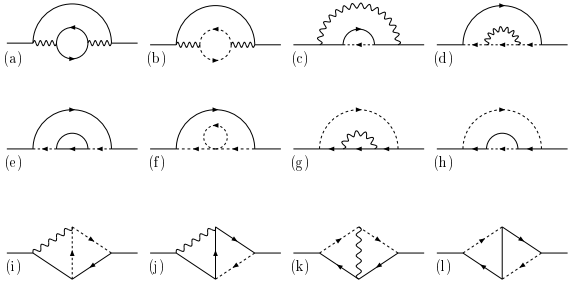}
%\end{center}
\caption{ %\baselineskip=23pt
Two-loop $O(\alpha_s^2)$ contributions to $\delta m_{\sg}^{(2,2)}$, 
with quark and squark propagators. Other diagrams obtained 
by charge conjugation are not shown. 
}
\label{fig2}
\end{figure}

The two-loop $O(\alpha_s^2)$ correction 
$\delta m_{\sg}^{(2)}$ consists of two parts, 
$\delta m_{\sg}^{(2)}=\delta m_{\sg}^{(2,1)}+\delta m_{\sg}^{(2,2)}$, 
where $\delta m_{\sg}^{(2,1)}$ is the contribution of the diagrams 
with only gluons and gluinos, while $\delta m_{\sg}^{(2,2)}$ is 
the remaining contribution including quark and squark loops. 
Two-loop diagrams for these contributions 
are shown in Fig.~\ref{fig1} and Fig.~\ref{fig2}, respectively. 
In these figures, 
the wavy line, solid line without an arrow, solid line with an arrow, and 
dashed line with an arrow represent the gluon, gluino, quark, and squark, 
respectively. 

The contribution $\delta m_{\sg}^{(2,1)}$ 
is obtained by applying the formula of the $O(\alpha_s^2)$ QCD 
correction to the quark masses \cite{avdeev} 
in the $\overline{\rm DR}$ scheme \cite{DR} to the SU(3) octet fermion. 
The result is  
\bea
&& \delta m_{\sg}^{(2,1)} = 
\nonumber\\ 
&& \left( \frac{C_V \alpha_s}{4\pi} \right)^2 M_3 
\left( -48\log\frac{M_3}{Q} + 36\log^2\frac{M_3}{Q} \right. 
\nonumber \\
&& \left. +26+5\pi^2 -4\pi^2\log 2 
+6\zeta_3 \right) , 
\label{YMpart} 
\eea
where $\zeta_3=\sum_{n=1}^{\infty}n^{-3}\simeq 1.202$. 
We have verified Eq. (\ref{YMpart}) by explicit calculation of the 
diagrams. At $Q=M_3$, the correction (\ref{YMpart}) is 
$\delta m_{\sq}^{(2,1)}/M_3\sim 31(\alpha_s/\pi)^2 \sim 0.03$. 

The contribution $\delta m_{\sg}^{(2,2)}$ including quark and squark 
loops is calculated by decomposition into 
two-loop basis integrals~\cite{tarasov,martin} using 
the integration by parts technique~\cite{tarasov,tarcer,grozin}, 
and their numerically evaluation by the package TSIL \cite{tsil}. 
We have analytically checked that the $Q$ 
dependence of $\delta m_{\sg}^{(2)}$ is consistent with 
the two-loop renormalization group equation \cite{mv1,rge} of $M_3$. 

The explicit form of $\delta m_{\sg}^{(2,2)}$ is rather long. 
Here we just show, for reference, the form in the 
limit of $m_{\sq}\gg M_3$:
\bea
\lefteqn{ \delta m_{\sg}^{(2,2)}(m_{\sq}\gg M_3) =} && 
\nonumber \\ 
&& 
\frac{\alpha_s^2M_3}{(4\pi)^2} 
\left[ 72 \log^2\frac{m_{\sq}}{Q} + 242 \log\frac{m_{\sq}}{Q} 
\right. \nonumber \\ && \left. 
+\log\frac{M_3}{Q}\left( 54 - 288 \log\frac{m_{\sq}}{Q} \right) 
-172 + \frac{14}{3}\pi^2 \right] 
\nonumber\\ 
&& +\frac{\alpha_s^2M_3}{(4\pi)^2} N_q C_VT_F 
\left( -8\log^2\frac{M_3}{Q} + \frac{52}{3}\log\frac{M_3}{Q}
\right. \nonumber \\ && \left. 
-\frac{37}{3}-\frac{4}{3}\pi^2  \right) . 
\label{asympt}
\eea
The last term of Eq.~(\ref{asympt}), which is independent of $m_{\sq}$, 
comes from the diagram (a) in Fig.~\ref{fig2}. 
We have checked that the $m_{\sq}$ dependence 
of Eq.~(\ref{asympt}) is consistent with 
the two-loop running of the gluino mass in the 
effective theory where squarks are integrated out \cite{split}. 

\section{Numerical results}

We present some numerical results of the $O(\alpha_s^2)$ 
pole mass of the gluino, for the running tree-level mass $M_3(M_3)=580$ GeV 
which is close to the values in the SPS1a point. 
The strong coupling constant within the standard model 
is given as $\alpha_s(m_Z)=0.12$. 

\begin{figure}[htbp]
%\begin{center}
\includegraphics[width=8cm]{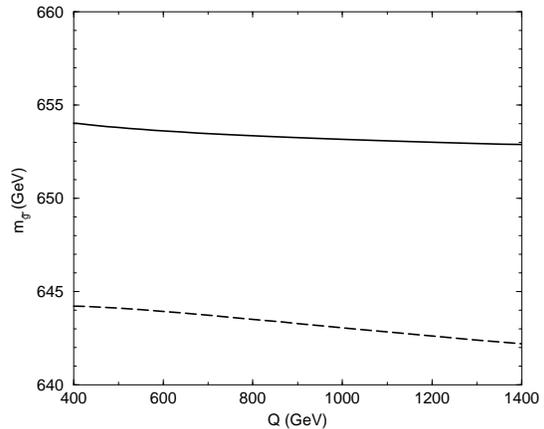}
%\end{center}
\caption{ %\baselineskip=23pt
Dependence of the one-loop (dashed) and 
two-loop (solid) pole masses of the gluino 
on the renormalization scale $Q$. 
Mass parameters are $(M_3,m_{\sq})=(580,800)$ GeV at $Q=580$ GeV. 
}
\label{figqren}
\end{figure}

We first show, in Fig.~\ref{figqren}, the residual dependence of 
the one-loop pole mass $m_{\sg}^{(1)}$ and 
two-loop pole mass $m_{\sg}^{(2)}$ on the renormalization 
scale $Q$, for the running squark mass $m_{\sq}(Q_0)=800$ GeV 
at $Q_0=580$ GeV. 
All parameters in the formulas are evolved by $O(\alpha_s^2)$ 
renormalization group equations. 
For reference, the tree-level $M_3(Q)$ decreases from 
589 GeV at $Q=400$ GeV to 559 GeV at $Q=1400$ GeV.  
%\begin{figure}[ht]
We see that the $Q$ dependence slightly improves by including 
$\delta m_{\sg}^{(2)}$. One should however note that, 
contrary to naive expectation, 
$\delta m_{\sg}^{(2)}$ is much larger than the $Q$-dependence of the 
one-loop result $m_{\sg}^{(1)}$. 

\begin{figure}[htb]
%\begin{figure}[htbp]
%\begin{center}
\includegraphics[width=8cm]{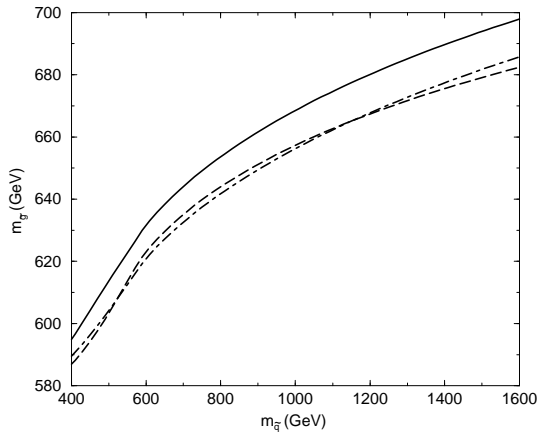}
%\end{center}
\caption{ %\baselineskip=23pt
The pole masses of the gluino at the one-loop with running masses (dashed), 
two-loop (solid), and modified one-loop mass with pole masses (dot-dashed), 
for the tree-level mass $M_3(M_3)=580$ GeV, as functions of $m_{\sq}(M_3)$. 
}
\label{fig4}
\end{figure}
In Fig.~\ref{fig4}, we compare $m_{\sg}^{(1)}$ and $m_{\sg}^{(2)}$ 
as functions of the running squark mass $m_{\sq}(Q=M_3)$. 
Here the renormalization scale is fixed at $Q=580$ GeV. 
The two-loop correction $\delta m_{\sg}^{(2)}$ is positive 
and in the range of $8-15$ GeV for $m_{\sq}=400-1600$ GeV. 
This correction is $O(1-2)$ \% of the one-loop result $m_{\sg}^{(1)}$, 
with partial cancellation between $\delta m_{\sg}^{(2,1)}$ 
and $\delta m_{\sg}^{(2,2)}$, but still 
similar to or larger than the expected uncertainty 
in the mass determination at future colliders \cite{msgdet1}. 
In Fig.~\ref{fig4}, we also show the modified one-loop 
result where the running masses in Eq.~(\ref{oneloop}) are 
replaced by the corresponding pole masses. This modification 
corresponds to the inclusion of higher-order corrections by one-loop 
renormalization group equations \cite{pierce2}. 
However, it is clearly seen that the resulting change of $m_{\sg}$ is 
much smaller than the two-loop correction $\delta m_{\sg}^{(2)}$. 

A recent work \cite{generaldmf} writes down the 
explicit form of the $O(\alpha_s^2)$ mass correction in terms of 
the basis integrals, including $m_q$ and left-right mixings of squarks. 
Since these parameters break the SU(2)$\times$U(1) gauge symmetry, 
their contributions to $m_{\sg}$ should be 
suppressed by factors $m_q^2/m_{\sq}^2$ or $m_q^2/m_{\sg}^2$ 
compared to the gauge symmetric contribution shown here. 
In addition, they only modifies contributions involving quarks and squarks 
in the third generation, while all generations contribute to 
the SU(2)$\times$U(1)-symmetric part of $\delta m_{\sg}^{(2,2)}$ 
with equal weight. 
We therefore do not expect that these SU(2)$\times$U(1)-breaking 
contributions are numerically relevant in 
future realistic studies of the SUSY particles. 
However, detailed study in cases of light gluino and/or squarks
is necessary for definite conclusion. 
%Even at the one-loop, it is observed \cite{pierce2} that 
%the SU(2)$\times$U(1)-breaking contribution is usually 
%below 2 \%, much smaller than 
%the SU(2)$\times$U(1) symmetric part. 

\section{Other two-loop corrections} 

\begin{figure}[htb]
%\begin{center}
\includegraphics[width=7.5cm]{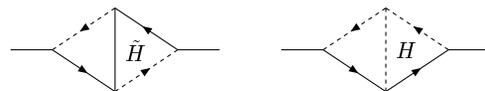}
%\end{center}
\caption{Some diagrams of the $O(\alpha_s h_q^2)$ contributions 
to the gluino pole mass.}
\label{diagram_y}
\end{figure}

Beyond one-loop, there are also mass corrections involving 
Yukawa and electroweak couplings. For example, 
diagrams involving Yukawa couplings $h_q$ of the 
Higgs bosons/higgsinos to quarks and squarks in the third generation, 
such as those in Fig.~\ref{diagram_y}, give the $O(\alpha_s h_q^2)$ mass 
corrections. 
Analytic form of these contributions can be 
derived from general formulas of the two-loop corrections 
to the fermion pole masses \cite{generaldmf} in the 
approximation of massless vector bosons in the loops. 
As a numerical example, 
again in the SU(2)$\times$U(1) symmetric approximation, 
the top Yukawa contribution for $M_3=m_{\sq}=m_{A^0}=\mu$, 
$\tan\beta=10$ is 
$\delta m_{\sg}^{(2,h_t)}/M_3\sim (2,10)\times\alpha_sh_t^2/(4\pi)^3$ 
for $A_t=(M_3,-M_3)$, respectively, which is much smaller 
than the $O(\alpha_s^2)$ contribution. 
We expect that the smallness of the Yukawa contribution 
would also hold in more general cases. 

\section{Conclusions}

We have calculated the two-loop SUSY QCD contribution 
to the gluino pole mass, 
ignoring SU(2)$\times$U(1) symmetry breakings in the loops. 
The $O(\alpha_s^2)$ correction to the gluino mass has 
been shown to be typically $1-2$ \%. 
For the case of $M_3(M_3)=580$ GeV, 
this correction is similar to, or larger than, 
the expected uncertainty in the mass determination from 
precision measurements at future colliders. 
The two-loop correction would be therefore important 
in the extraction of $M_3$ from experimental data and 
the determination of the SUSY breaking at the unification scale.

\end{document}